\newcommand{\lis}{\mbox {$ l_{1}^{\sigma}$}}
\newcommand{\qir}{\mbox {$ q_{1}^{\rho}$}}

\newcommand{\pin}{\mbox {$ p_{1}^{\nu}$}}
\newcommand{\al}{\mbox{$\alpha $}}

\newcommand{\s}{\mbox{$\sigma $}}

\newcommand{\mi}{\mbox{$\mu _{1}$}}
\newcommand{\mt}{\mbox{$\mu _{2}$}}
\newcommand{\mn}{\mbox{$\mu _{n}$}}
\newcommand{\mnm}{\mbox{$\mu _{n-1}$}}
\newcommand{\mnp}{\mbox{$\mu _{n+1}$}}

\newcommand{\e}{\mbox{$e^{ik_{0}X}$}}
\newcommand{\qe}{\mbox{$e^{iq_{0}X}$}}

\documentstyle[12pt]{article}
\newcommand{\kim}{\mbox {$ k_{1}^{\mu}$}}

\newcommand{\ki}{\mbox {$ k_{1}$}}

\newcommand{\qi}{\mbox {$ q_{1}$}}

\newcommand{\qo}{\mbox {$ q_{0}$}}
\newcommand{\ko}{\mbox {$ k_{0}$}}

\newcommand{\be}{\begin{equation}}
\newcommand{\br}{\begin{eqnarray}}
\newcommand{\ee}{\end{equation}}
\newcommand{\er}{\end{eqnarray}}

\newcommand{\ddS}{\mbox {$\frac{\delta}{\delta \Sigma}$}}

\newcommand{\p}{\mbox {$ \partial$}}

\newcommand{\pp}{\mbox {$ \partial ^{2}$}}
\newcommand{\mup}{\mbox {$ \partial _{\mu}$}}
\newcommand{\nup}{\mbox {$ \partial _{\nu}$}}
\newcommand{\Xmi}{\mbox {$ X^{\mu _{1}}$}}
\newcommand{\Xmt}{\mbox {$ X^{\mu _{2}}$}}
\newcommand{\Xmnm}{\mbox {$ X^{\mu _{n -1}}$}}
\newcommand{\Xmn}{\mbox {$ X^{\mu _{n}}$}}
\newcommand{\Xmnp}{\mbox {$ X^{\mu _{n +1}}$}}

\begin{document}
\title{Proper Time Formalism, Gauge Invariance and the Effects of
a Finite World Sheet Cutoff in String Theory}
\author{B. Sathiapalan\\ {\em
Physics Department}\\{\em Penn State University}\\{\em 120
Ridge View Drive}\\{\em Dunmore, PA 18512}}
\maketitle
\begin{abstract}
We discuss the issue of going off-shell in the proper time
formalism.  This is done by keeping a finite world sheet
cutoff.  We construct one example of an off-shell covariant
Klein Gordon type interaction.  For a suitable choice of the
gauge transformation of the scalar field, gauge invariance is
maintained off mass shell.  However at second order in the
gauge field interaction, one finds that (U(1)) gauge invariance
is violated due to the finite cutoff.  Interestingly, we find,
 to lowest order,
that by adding a massive mode
with appropriate gauge transformation laws
to the sigma model background,
one can restore
gauge invariance.
The gauge transformation law is found
to be consistent, to the order calculated, with what one expects
from the interacting equation of motion of the massive field.
We also extend some previous discussion on applying
the proper time formalism for
propagating
gauge particles, to the interacting (i.e. Yang Mills)
case.
\end{abstract}
\newpage
\section{Introduction}

  The sigma-model or renormalization group approach to string theory
[1-7] has shown some promise as an
alternative to string field theory for doing
non-trivial calculations.  It is hoped that it will be computationally
simpler and that
the physical significance of the symmetries be more transparent
than in string field theory [8-12].
In the renormalization group approach to string
theory there are two outstanding issues.  One is that of gauge invariance
and the other is that of an off-shell formulation.  These two issues are,
of course, intertwined because, in general, maintaining gauge invariance
off-shell is more difficult than when on shell.  The issue of `massless'
U(1) gauge invariance in the on-shell sigma model formalism has been
discussed some time ago [13,28-30].  Gauge invariance
associated with the massive
modes is a little more complicated and requires the introduction of an
infinite number of `proper time' variables and was discussed in ref[14].
The discussion of gauge invariance in these papers has been restricted to
linear gauge transformations determined by the
invariance of the free theory.
At the interacting level the situation is a lot more complicated,
especially off-shell.  A discussion of these issues
in the BRST formalism is contained in Ref[27].

In ref[15] a study of the gauge invariance of the interacting theory
was initiated.  We derived, in the proper time formalism [16,21-26]
(which
is really a variant of the renormalization group approach), the covariant
Klein Gordon equation.  It was shown that the technique works equally
well for point particles as well as strings.  For point particles
it is exact, while for strings it is derived as a
low energy approximation.
The usual gauge invariance at the massless level,
$ \delta A_{\mu} = \mup \Lambda$,
arises as a freedom to add total derivatives to the two dimensional
world sheet action.  It was shown that this is no longer a symmetry
when interactions are present because of boundary terms.  These
boundary terms can be cancelled by an appropriate transformation of the
Klein Gordon scalar field $\delta \phi = i \Lambda \phi $.  This
enables us to understand how the `interacting'
terms in gauge transformations
arise in this formalism. (By `interacting terms' we mean those that
are required specifically by the interacting theory).  We had also
discussed a possible generalization of the proper time formalism when
dealing with the propagation of gauge particles.  As a first application,
we gave (yet another!) derivation of the Maxwell equations.

   We would like to extend the results of that work in two directions.
One is to study the effect of keeping a finite cutoff on the world sheet.
The motivation for this goes back to Ref[17], where it was argued that
in order to go off-shell, in this formalism, one needs a finite cutoff.
In the language of the renormalization group this is equivalent to the
statement that when the cutoff is finite (w.r.t. the correlation length)
one is away from the fixed point (i.e. off-shell) and this is also where
the irrelevant operators (i.e. vertex operators for off-shell
massive modes) are
no longer irrelevant.  Thus, in this paper
we construct an off shell vertex coupling the
photon to two scalars.
There is an obvious consistency check.  The equation
of motion one derives should be an off-shell version of a gauge covariant
Klein Gordon equation, and should reduce to the usual one on
shell.\footnote { Here we use the word `off-shell' to describe fields that
do not satisfy the momentum mass shell condition $p^{2} = m^{2}$.  They
may or may not satisfy the full equations of motion.} We should point
point out that the vertex constructed is not unique - there
are many other
possibilities.  Other constraints have to be imposed to single out
one choice.

 We then proceed
to the next order, which is the cubic term in the covariant Klein Gordon
equation - involving two vector fields.  Interestingly enough, we find
that one can restore gauge invariance (at least to lowest order)
by adding
a massive ``spin-2''
particle with an appropriate gauge transformation law.
Thus, gauge invariance can be maintained with finite cutoff if we
have nontrivial backgrounds for the massive modes.  We can also
check whether
the gauge transformation law for this massive mode is consistent with its
(interacting) equation of motion.  We find that this is so, at
least, to lowest
order in momentum.  Assuming these results continue to hold to all orders
in the cutoff, which will require adding an infinite tower of
 massive modes, we can
say,
 that, in a sense, keeping a non zero world sheet cutoff is equivalent
to keeping all the massive modes.
We do not find this statement surprising.
Given the renormalization group interpretation, it is to be expected.
Nevertheless, we find it very interesting that it is being derived
in a completely different way, with no reference to the renormalization
group, whatsoevser.
In ref[14], however, we speculated that this is true for a non zero
space time
cutoff (rather than just the world sheet cutoff). But we have no
calculations
to support this speculation, yet.

We also extend the results of ref[15] in another direction.  We study
the Yang-Mills vector field and use the generalization of the proper-time
method to derive its equation of motion (to lowest order in momentum).
This is only a vindication of the method described there for
gauge particles,
since the result is standard.  What would be non trivial is to
extend this to
a finite cutoff off shell vertex.

This paper is organized as follows:  In Section II we present a brief
review of the derivation of Maxwell's equation and then derive the
Yang Mills case.  In Section III we discuss the off shell three point
vertex of the covariantised Klein Gordon equation.  In Section IV we look
at the four point vertex and study the gauge covariance properties.  We
conclude in Section V.  Details of a calculation is given in an Appendix.

\newpage
\section{Equation of Motion for Gauge Fields}
\setcounter{equation}{0}

The renormalization group equation can be rephrased as
a `proper time' equation
 of the form:
\be
(\frac{d}{dln(z-w)}+2)<O(z)O(w)>=0
\ee
where $O(z)$is a vertex operator.

This equation says that $O(z)$ has dimension one.  Since
$z= e^{\tau + i \sigma}$
where, for open string vertex operators, we have to set $\s = 0$,
this equation
takes the form of a proper time equation familiar for point particles.  In
Ref[15] we used this to derive the covariant Klein Gordon equation.
 The expectation value was evaluated with a weight
 $e^{\frac{1}{2} \int d^{2}z \partial
 _{z} X \partial _{\bar{z}} X + \int dz A_{\mu}
\partial _{z} X^{\mu} }$ where $A_{\mu}$ is a background gauge field and
$O(z) = \phi [X(z)]$ is the scalar field.  We refer the reader to
Ref[15] for the details.  If instead of $\phi [X(z)]$ we want to
consider a gauge field, then $O(z) = A_{\mu} [X(z)] \p _{z} X ^{\mu}$.
However this naive substitution is not satisfactory since we need to
ensure that the result is invariant under $A_{\mu} \rightarrow A_{\mu}
+ \mup \Lambda$.  The solution proposed in ref[15] is to replace
$O(z)$ by $\int dz O(z) = \int dz A_{\mu} \p _{z} X ^{\mu}$.  This is
invariant, because under a gauge transformation  $A \p X$ changes by a
total derivative which vanishes when integrated - asssuming that there
are no boundaries.  However, in this case the operator $\frac{d}
{d ln (z-w)}$ is replaced by the functional derivative $\frac{\delta}
{\delta \Sigma (z-w)}$ where $\Sigma (z-w)= <X(z)X(w)>$.
Thus $\Sigma (z-w)$ is treated as a {\em field}, rather than as a
function.\footnote{This is very similar to evaluating
derivatives w.r.t. the Liouville mode - something that has
been used frequently
 to derive equations of motion [12,13,14].}  Thus we have the operation
\be
\int dz \int dw \ddS <O(z) O(w)> = 0
\ee
\be
\Rightarrow \int dz \int dw \ddS <A_{\nu} \p _{z} X^{\nu} A_{\mu} \p _{w}
X^{\mu} > =0
\ee
with $A_{\mu} (X) = \int dk A_{\mu}(k) e^{ikX}$.  The coefficient
of $A_{\nu} $ gives the equation of motion.  To obtain
the linear term in the equation of motion one does not need any
insertions of $\int dz A_{\mu} \p _{z} X^{\mu}$ from the sigma
model action.  So we get:
\be
\int dz \int dw \ddS [A(p).kA(k).p \p _{w} \Sigma
\p _{z} \Sigma + p.kA(p).A(k) \Sigma \p _{z} \p _{w} \Sigma]
=0
\ee
\be \label{eq:II.5}
\Rightarrow [-A(p).kA(k).p + p.kA(p).A(k)]\p _{z} \p _{w} \Sigma =0
\ee
The coefficient of $A_{\mu}(p)$ is thus
\be
-k^{\mu}A(k).p + p.k A^{\mu}(k) = 0   \label{eq:II.6}
\ee
Thus we get Maxwell's equation.
Note, also, that the expression in square brackets in (\ref{eq:II.5})
is in fact the Maxwell action, (from which (\ref{eq:II.6}) is obtained
by varying w.r.t $A_{\mu}$) except for a factor of 1/2.
This factor of 1/2 is important in the Yang Mills case where
we are concerned with the relative normalization vis-a-vis the cubic
and quartic terms in the action.  Thus, the proper time formalism
gives the equations of motion of Yang Mills theory (as we shall see),
and not the action.
To reconstruct the action, one will have to put in by hand these
factors of 1/2 , 1/3 or 1/4
(in the cubic and quartic terms respectively).

The crucial point to note is that we
are able to integrate by parts on the variable `$w$'
since we have an integral
$\int dw$.  This concludes our review.

We now turn to the Yang Mills case where we would like
to get the cubic and quartic pieces
in a manifestly Lorentz covariant fashion.  This calculation
can be simplified by the following identity (proved in the Appendix)
\be
\e i \ki \p X = \int _{0}^{1} d \al \p (e^{i\alpha \ko X} i \ki
X) + \int _{0}^{1} d \al [ e^{i\alpha \ko  X} X^{\nu} \p X^{\mu}
(i)^{2} \al \ko ^{[\nu} \ki ^{\mu ]} ]   \label{eq:Id}
\ee
We have written the LHS as a sum of two terms, one of which, involving
$F_{\sigma \mu} = \p _{\sigma} A_{\mu} - \mup A_{\sigma}$
,is manifestly gauge invariant (under $A_{\mu} \rightarrow  A_{\mu} +
\mup \Lambda $) and the other one is a total derivative that vanishes
if there are no boundaries.  Furthermore the first term on the RHS
can be written as
\be
\int ^{1}_{0} \frac{d \al }{\al} \p (e^{i \alpha (\ko + \ki ) X})
\ee
if we remember to extract the piece that is linear in $\ki$.

We can map the upper half plane to a circular disc so that the vertex
operators are attached to the circular boundary.

To calculate the cubic term, we have to bring down one factor of
$\int A \p X$ from the exponent.
Thus we have to consider
\be    \label{eq:II.7}
\oint dz \oint du < i A_{\mu} \p _{z}X^{\mu} \int _{u}^{z} dw
i A_{\nu}\p _{w} X^{\nu} i A_{\rho} \p _{u} X^{\rho}>
\ee
Here $A_{\mu} \equiv A_{\mu}^{a} T^{a}$, where $T^{a}$ satisfy
the Lie Algebra $[T^{a}, T^{b}] =i f^{abc}T^{c}$ and $Tr(T^{a} T^{b})
= \delta ^{ab}$.  In correlations we will assume that there is a trace
over the matrix indices.  Using the notation of ref[14] we will
write $A^{\mu}(X) = \int dk_{0} dk_{1} k_{1}^{\mu} \Phi (k_{0},k_{1})
\e$ and for simplicity we will omit the integrals and the field $\Phi$
when we write correlations.
Thus (\ref{eq:II.7}) becomes :
\be     \label{eq:II.8}
Tr(T^{a}T^{b}T^{c})\oint dz \oint du \frac{\delta}{\delta \Sigma}
<i\kim  \p _{z} X ^{\mu}\e i\int _{u} ^{z} dw
p _{1} ^{\nu} \p _{w} X ^{\nu} e^{ip_{0}X} i \qi ^{\rho} \p _{u}
X^{\rho} \qe >
\ee
In performing the above calculation we must keep in mind that
there is a path ordering, implicitly, for the matrices.  Thus the
ordering of the matrices inside the trace follows the ordering
of the three points $u, w $ and $z$ along the circle.  Since the
vector bosons are bosons and the resultant interaction is
symmetric under permutations, we can restrict ourselves
to a particular ordering while evaluating expressions like
(\ref{eq:II.8}), and mutiply the result by a
combinatoric factor, which is equal to the number of
permutations.  In the cubic case there is only one insertion,
so there are no such factors.

We now substitute the RHS of (\ref{eq:Id}) for each of the three
factors inside the correlation in (\ref{eq:II.8}).  Amongst the
many terms that arise, is the following one:
\be             \nonumber
\int _{0}^{1} d \al
\int _{0}^{1}
d \beta \int _{0}^{1}
d \gamma \gamma
\oint dz \oint du
\ee
\be         \label{eq:II.9}
<i \kim \p _{z} (X ^{\mu} e ^{i \alpha k_{0} X}) \int _{u}^{z} dw
i p_{1} ^{\nu} \p _{w} ( X^{\nu} e^{i \beta p_{0}X}) i^{2}
\qo ^{[\rho} \qi ^{\sigma ]} X^{\rho} \p _{u} X^{\sigma}
e^{i\gamma q_{0}X}>
\ee
There are also two other terms of this type obtained by
interchanging $k \leftrightarrow q$ and $p \leftrightarrow q$.
(\ref{eq:II.9}) now becomes (the $\alpha , \beta $and$ \gamma
$ integrals are understood):
\be                       \label{eq:II.10}
1/2\oint dz \oint du \kim \pin < \p _{z} (X^{\mu}e^{i \alpha
k_{0} X }) [ (X^{\nu} e^{i \beta p_{0}X(z)}) - (X^{\nu}
e^{i\beta p_{0} X(u)})] X^{\rho} \p _{u} X^{\sigma} e ^{i
\gamma q_{0} X} > \qo ^{[\rho} \qi ^{\sigma ]}
\ee
We are interested in the dependence on $\Sigma (z-u)$
in the above expression. \footnote {We ignore the logarithmically
divergent terms.  The justification for this is that (see
\cite{BS3}) in the final answer $ln a $ and $ln (z-u)$
will always occur in the form $ln( \frac{z-u}{a} )$, purely
for dimensional reasons.  Thus we can concentrate on the
$ln (z-u)$ terms, and set $ln a$ , formally, to zero.}

To obtain the cubic coupling of Yang-Mills theory we only
need to keep the lowest order terms in (\ref{eq:II.10}).
In particular the second term inside the square brackets does
not contribute anything, to this order.  We can also ignore,
to this order,
the exponential factors.
Thus we get ($\Sigma (z-u) \equiv \Sigma $):
\be      \label{eq:II.11}
1/2 \oint dz \oint du \kim \pin \qo ^{[\rho} \qi^{\sigma ]}
( \delta ^{\mu \rho} \delta ^{\nu \sigma} \p _{z} \Sigma
\p_{u} \Sigma + \delta ^{\mu \sigma} \delta ^{\nu \rho}
\Sigma \p _{z} \p _{u} \Sigma )
\ee
Restoring all other factors from (\ref{eq:II.7}),(\ref{eq:II.8}) we get
\be       \label{eq:II.12}
1/2 A_{\mu}^{a} (k_{0}) A_{\nu}^{b} (p_{0}) \qo _{[\rho}
A_{\sigma ]}^{c} (\qo )1/2 f^{abc} \oint dz \oint du
( \delta ^{\mu \rho} \delta ^{\nu \sigma}
- \delta ^{\mu \sigma} \delta ^{\nu \rho} ) \frac{\delta}{\delta
\Sigma} (\Sigma \p _{z} \p _{u} \Sigma )
\ee
Note that we have integrated by parts on $z$.
In evaluating the matrix trace, we have used antisymmetry in
`$a$' and `$b$'.  The coefficient of $A_{\mu}(k_{0})$ gives the
quadratic piece of the Yang-Mills equation of motion.
The full symmetry between $k, p, $ and $q$
is manifest when we include the two other terms
from eq(\ref{eq:II.8}) mentioned earlier.  As in the case of the
free Maxwell case discussed earlier, we see the full cubic term
of the Yang-Mills {\em action} in (\ref{eq:II.12}) (we also have to
add the two other terms necessary for symmetry).  Again, as before
we have to include a factor of 1/3 to get the right normalization.

The quartic term can similarly be obtained by calculating:
\be       \label{eq:II.13}
\frac{2!}{2!} \oint dz \oint du \frac{\delta}{\delta \Sigma}
<A \p _{z} X \int _{u} ^{z} dw A \p _{w} X
\int _{u}^{w} dv A \p _{v}X  \,
A\p _{u} X >
\ee
The 2! in the denominator comes from expanding the exponential
and the 2! in the numerator comes from having chosen a particular
ordering , $w > v$ in (\ref{eq:II.13}).  In the notation of
ref[14] this becomes:
\be    \label{eq:II.15}
Tr(T^{a}T^{b}T^{c}T^{d})\oint dz \oint du < \kim \p _{z}
X^{\mu} \e \int_{u}^{z} dw \pin \p _{w} X^{\nu} e^{i p_{0} X}
\int _{u}^{w} dv \qir \p _{v} X^{\rho} \qe \,
\lis \p _{u} X^{\sigma}
e^{i l_{0}X} >
\ee
Since we are only interested in the quartic Yang-Mills piece
we can set all the momenta to zero.  In that case we get
on doing the `$v$' integral:
\be        \label{eq:II.16}
\oint dz \oint du <\kim \p_{z} X^{\mu} \int _{u}^{z} dw \pin
\p_{w} X^{\nu} \qir [X^{\rho} (w) - X^{\rho}(u)] \lis \p_{u}
X^{\sigma}>
\ee
Consider the term involving $X^{\rho}(w)$.  We can rewrite the
region of integration (and still preserve the ordering)
as $\oint dw \oint du \int _{w}^{u} dz$.
This gives
\be     \label{eq:II.17}
\oint dw \oint du < \lis \p_{u} X^{\sigma} \int _{w}^{u} dz
\kim \p _{z} X^{\mu} \pin \p_{w} X^{\nu} \, \qir X^{\rho}(w)>
\ee
Performing the $z$ integral and evaluating the correlation gives:
\be        \label{eq:II.18}
\oint du \oint dw \lis \kim \pin \qir
(\delta ^{\sigma \nu} \delta ^{\mu \rho} \Sigma \p _{u} \p _{w}
\Sigma +
\delta ^{\sigma \rho} \delta ^{\mu \nu} \p _{u}\Sigma  \p _{w}
\Sigma )
\ee
Since we can integrate by parts on $u$ or $w$, we can see that
(\ref{eq:II.18}) is antisymmetric in $\sigma , \mu$ and
$\nu , \rho $ indices respectively.
Antisymmetry in $\sigma , \mu$ indices implies an antisymmetry
in $a,d$ indices.  Thus, restoring the group theory factors,
we get:
\be          \label{eq:II.19}
1/4 f^{dae} f^{bce} 1/4 l^{[ \sigma} \kim ^{]} p_{1}^{[\nu}
\qi ^{\rho ]}
(\delta ^{\sigma \nu} \delta ^{\mu \rho} \Sigma \p _{u} \p _{w}
\Sigma +
\delta ^{\sigma \rho} \delta ^{\mu \nu} \p _{u}\Sigma  \p _{w}
\Sigma )
\ee
The second term in (\ref{eq:II.16}) gives a permutation of this.
Thus we get
(rexpressing in terms of $A$):
\be            \label{eq:II.20}
\oint du \oint dw [1/2 f^{ade} f^{bce}
A^{a}_{\mu}(k_{0})
A^{b}_{\nu}(p_{0})
A^{c}_{\rho}(q_{0})
A^{d}_{\sigma}(l_{0})
(\delta ^{\sigma \nu} \delta ^{\mu \rho}  -
\delta ^{\sigma \rho} \delta ^{\mu \nu})]\frac{\delta}{\delta \Sigma}
\Sigma  \p _{u} \p _{w}
\Sigma .
\ee
As before, the coefficient of $A(p_{0})$ gives the
contribution to the equation of motion, and we can recognize
 the Yang-Mills structure (after appropriate symmetrization)
 .  Also as before, we can
recognize in the square brackets the quartic term  of the Yang-Mills
{\em action}.  As mentioned earlier, in going from the equation
of motion to the action, we have to divide by 4 in order
to get the right normalization.
The action that gives the above equation of
motion is
\be
1/8 [\p_{[\mu} A_{\nu ]}^{c} + f^{abc} A_{\mu} ^{a}A_{\nu} ^{b}]
^{2}
\ee
The rest of the terms in (\ref{eq:II.8}) and (\ref{eq:II.15})
represent higher order string corrections to the above action.

Thus we have demonstrated a method for dealing with interacting gauge
particles in the framework of the proper time formalism.  The crucial
point is to treat the propagator $<X(z) X(w)>$ as a field ($\Sigma$),
and keep the integrals over the coordinates $z$ and $w$.  This
allows one to integrate by parts when performing functional
differentiation w.r.t $\Sigma$.  In this paper we have stayed close
to the mass-shell.  To go off mass-shell in a manifestly gauge
covariant way requires a further extension of these methods.
In the next section we will address this problem for the simpler
 case of the Klein-Gordon equation.
\newpage
\section{Going Off-Shell with a Finite Cutoff}
\setcounter{equation}{0}

If one evaluates
\be
(\frac{d}{dln(z-w)} +2)<\phi [X(z)] \phi [X(0)]> \label{eq:III.1}
\ee
with the interaction $\int A_{\mu} \p _{z} X ^{\mu} dz $
in the sigma model action,
one gets an equation of motion for $\phi $ that has arbitrarily
high powers of $A_{\mu}$.  There are many ways of understanding this.
In terms of the renormalization group, $ \int A_{\mu} \p _{z} X^{\mu} dz$
is a marginal operator (at least for small momenta).  In the continuum
limit, near a fixed point, the $\beta $-function can have terms with
arbitrarily large powers of the (marginal) coupling constant and these
terms have no suppression factors on dimensional grounds.
In terms of Feynman diagrams in string
theory, one is looking at tree diagrams with external massless fields,
of which, there can be any number.  In terms of the equations of
motion, the
massive fields can be solved for in terms of massless fields,
and can be eliminated from the equations.

If one wants to be sensitive to the coefficients
of the irrelevant operators and not just the
marginal ones, then one should look at distance scales
on the order of the under
lying lattice cutoff.  In this
situation the $\beta $-functions are finite degree polynomials and
involve all the coupling constants.
This is the case when one is far from the
fixed point.  In string theory terms, one is off shell and the massive
modes are important and the equations
of motion involve all the fields, not
just the massless ones.  The equations
are expected to be polynomial in such
a situation.  This is the situation described by the cubic string field
theory vertex [11].

Thus at the moment we have two extreme situations-
the quadratic equations of string field
theory and the infinitely non polynomial
equations of the sigma model approach.
It seems plausible that one should be able to interpolate between these
two extremes.  The renormalization group interpretation suggests that by
varying the value of the underlying cutoff one can modify the degree to
which heavy fields are ``integrated out''.  In ref[16] we showed
how this could be done in the proper time formalism:  When
evaluating eqn. (\ref{eq:III.1})
we introduce a lattice spacing `$a$' and require that
it be the distance of closest approach between two vertex operators.
The parameter $z/a$ determines how many vertex operators can be
inserted between $\phi [X(z)] $ and $ \phi [X(0)]$ and thus the degree
of polynomiality of the equation.  In particular, if $z=2a$ one can
insert only one vertex operator and the equations are purely quadratic.
If $z/a \rightarrow \infty $ we get a polynomial of arbitrary high
order - the sigma model situation.  Thus for different values of
$z/a$ one gets different sets of equations.  One expects that if any
of these sets of equations is expressed completely in terms of massless
fields, by eliminating the massive ones, we will end up with the
equation obtained in the $z/a \rightarrow \infty$ case, i.e. the sigma
model case.
However we do not yet have a proof of this.
We also do not have field theory
Feynaman rules for calculating higher n-point
functions from a given set of
lower n-point vertices and propagators.  Without such a prescription
we cannot really check the consistency of our formulation.  This is an
important issue that we hope to address in the future.
Meanwhile, however,
there is one basic consistency check that can be done, and that is
to check gauge invariance.  We can require that our equation of motion
be gauge covariant and that it reduce to the covariant Klein Gordon
equation on-shell.

Before we do this, let us note that in order
to make contact with critical strings one
should be careful about group theory factors.  However, in
this work we will just discuss a charged string in some background
U(1) field.  The problem of a charged string moving in an
electromagnetic background has been discussed in ref[28].  The
point of deviation in our discussion is that we will be keeping
a finite cutoff in order to go off-shell.  Thus the electromagnetic
field can have any momentum dependence.

Thus, following ref[16], we consider the proper time equation
\be                     \label{eq:III.2}
z\frac{d}{dz}\left\{ z^{2} <e^{ik'X(z)} \int _{a}^{z-a} du
A_{\mu} [X(u)]\p _{u} X^{\mu} (u) e^{ikX(0)}>\right\} \phi (k)
=0
\ee
This is the same as eqn (3.1) (with one insertion of $A$)
where $\phi [X(z)]$ has been
chosen to be $e^{ik'X(z)}$. The factor $z^{2}$ in eqn.(3.2)
is needed to produce the second term in eqn.(3.1).
Before we proceed to evaluate (\ref{eq:III.2})
let us evaluate a simpler
quantity, namely, the gauge transformation of (\ref{eq:III.2})
under
$A_{\mu} \rightarrow  A_{\mu} + \mup \Lambda$.
If we replace $A_{\mu}$ by
$ A_{\mu} + \mup \Lambda$
in the covariant Klein-Gordon equation
we get $2 \mup \Lambda \mup \phi + \pp \Lambda \phi $ to lowest order.
We would first like to make sure that we reproduce this.
We get:
\be     \label{eq:III.2.1}
z\frac{d}{dz}\left\{ z^{2} <e^{ik'X(z)} [\Lambda (q) e^{iqX(z-a)}
-\Lambda (q)e^{iqX(a)}]
 e^{ikX(0)}>\right\} \phi (k)
=0
\ee
where we have gone over to the momentum representation for $A $ and
$\Lambda $.
\be
z\frac{d}{dz}[
a^{k'.q}z^{k'.k+2}(z-a)^{q.k}
-(z-a)^{k'.q} z^{k'.k+2}a^{q.k}
]\phi (k) =0
\ee
This gives
\be
[(k'.k+2)
a^{k'.q}z^{k'.k+2}(z-a)^{q.k}
+q.k a^{k'.q}z^{k'.k+3}(z-a)^{q.k-1}
\ee
\[
-(k'.k+2)(z-a)^{k'.q} z^{k'.k+2}a^{q.k}
-k'.q (z-a)^{k'.q-1} z^{k'.k+3}a^{q.k}
]\phi (k) = 0
\]
On-shell, setting $k.k' +2=q.k =q.k'=0$ in the exponents, we get the
leading order \footnote{
Upto overall powers of $a$, which can be gotten rid of by
renormalizing the vertex operators and using momentum conservation
: each vertex operator comes with a factor
$a$ raised to the scaling dimension of the operator, which in turn is
equal to $k^{2} /2 $+ number of derivatives in the vertex operator.}
contribution:
\be
(q.k-q.k')\frac{z}{z-a} \Lambda (q) \phi (k) =
(q^{2} + 2q.k)\Lambda (q) \phi (k) \frac {z}{z-a} =0
\ee
where we have used $k+k'+q=0$.  What it should give on-shell is,
of course,
$(q^{2} + 2 q. k )\Lambda (q) \phi (k) =0$, which is just the
change, under $ \delta \phi = i \Lambda \phi $, of the Klein Gordon
equation.  We can get rid of the factor $\frac{z}{z-a}$
by replacing the
factor $z^{2}$ in eq. (\ref{eq:III.2})
by $z(z-a)$.  We will do this from now on.
In the limit $a \rightarrow 0$, this does not make any difference.
But for finite $a$, the proper time equation is modified.
Thus, we evaluate (\ref{eq:III.2})
 with $A_{\mu}[X(u)] = \int dq A_{\mu}(q)
e^{iq.X(u)}$ and $z^{2}$ replaced by $z(z-a)$ and find the result
\be    \label{eq:III.3}
\{(A.k + \frac{A.k'}{q.k'})(x-1)^{q.k}x^{k'.k+2} -
\ee
\[
(A.k'+\frac{A.k'}{q.k'})(x-1)^{q.k'}x^{k'.k +2} +
\]
\[
  (k'.k+1) \frac{A.k'}{q.k'}(x-1)^{q.k+1}x^{k'.k+1}  -
\]
\[
  (k.k'+1)\frac{A.k'}{q.k'}(x-1)^{q.k'+1}x^{k'.k +1} +
\]
\begin{eqnarray*}
 (x-1)(q^{2}A^{\mu}-q.Aq^{\mu})k_{\mu} \int _{1}^{x-1} du
(x-u)^{k'.q-1}u^{q.k-1} x^{k'.k+2} +            & \\
\frac{(q^{2}A^{\mu}-q.Aq^{\mu})}{q.k'}k_{\mu}[(k.k'+1)(x-1) +x]
\int _{1}^{x-1}du (x-u)^{k'.q} u^{q.k-1} x^{k'.k+1} & \}\phi (k) =0
\end{eqnarray*}
The last two terms vanish when the photon is on-shell(i.e. $\mup F_{\mu
\nu} = 0$).
We also note that the appearance of a pole at $k'.q =0$ is deceptive
, since the pole terms cancel out.  If we choose $x=2$,
 the integrals vanish,
and we get
\be
(A.k-A.k')2^{k'.k+2}\phi (k) = (2A.k+A.q)2^{-k^{2}-k.q+2}\phi (k)=0
\ee
Furthermore, on-shell, if we set $k'.k +2=q.k=q.k'=0$, (3.7) reduces to
\be
(A.k-A.k')\phi (k) = (2A.k+A.q) \phi (k) =0
\ee
In this case we can also let $a \rightarrow 0 $ or $x \rightarrow \infty $.
Thus, the main result of this section is the contribution to the
Klein Gordon equation given by (\ref{eq:III.3})
The integrals can be expressed in
terms of hypergeometric functions, but we will not do so here.

On the face of it expression (\ref{eq:III.3})
 looks like yet another 3-pt. vertex
that should be obtainable from some string field theory [11,12].
However, when
we study (\ref{eq:III.2})
 and (\ref{eq:III.3}) we see a difference.  In
(\ref{eq:III.2} there is an integral over $u$, the location of $A_{\mu}$.
The vertices
considered in the literature thus far always have the three vertex
operators in specific locations.  In the special case of $z=2a$  these
integrals (in eqn. (\ref{eq:III.3})) vanish, and we
have well defined locations for
the vertex operators.

We also have a rule
that `$a$' is the minimum
spacing between two vertex operators.  Thus, for
instance, when we consider the gauge transformed kinetic term
in the equation of
 motion, we get
\be                              \label{eq:III.4}
z\frac{d}{dz}[z(z-a) <\phi [X(z)] (\Lambda [X(z-a)] -
\Lambda [X(a)] )\phi [X(0)] >] =0
\ee
The important point is that
\be     \label{eq:III.5}
\delta \phi [X(0)] = i \Lambda [X(a)] \phi [X(0)]
\ee
and not
\be
\delta \phi [X(0)] = i \Lambda [X(0)] \phi [X(0)]
\ee
In fact, we have seen that when we substitute $A_{\mu} \rightarrow
A_{\mu} + \mup \Lambda $
we get eqn. (\ref{eq:III.2.1}) (with $z^{2}$ replaced by $z(z-a)$),
which is identical with (\ref{eq:III.4})
.  Thus it is
the transformation
 law (\ref{eq:III.5})
that is consistent with the gauge invariance of
(\ref{eq:III.3}).
However, until we have a full formulation that spells out
the precise relation between the (N+1)-point function and the N-point
function we cannot claim that this rule of keeping a minimum spacing
`$a$' between vertex operators is consistent.

         If we choose $z=2a$, then the equation of motion is
         quadratic in the fields.  In this case the variation
$\delta \phi = -i \Lambda \phi $
of the $A \phi $ piece has to be cancelled
by the variation of a massive mode rather than by the variation of the
$A-A-\phi $ piece as in point particle field theory.  If $z/a > 3$
one can have two or more powers of A.  In this case one expects
a cubic $A-A- \phi $ term in the equation of motion.
The gauge transformation property
of this term is the topic of the next section.

There are also other modifications that are possible.
  One can imagine doing the
 same calculation on a disc
where cyclic symmetry is manifest.
This would be similar to what is done in ref[18,19,20].  In this case
the equation of motion would be different.

Our main aim in this paper is to
explore the issues that arise in keeping a finite cutoff - particularly
issues of gauge invariance.  We should keep in mind that geometries
other than that used in eq (\ref{eq:III.3})
are possible.  The results of ref[11,12],
in fact, suggest that manifest cyclic symmetry is very important
for the full gauge invariance of the theory.
\newpage
\section{Gauge Invariance and Finite Cutoff}
\setcounter{equation}{0}

We have been using gauge invariance as a consistency
check on the off-shell
terms in the Klein Gordon equation.  At the next order in $A_{\mu}$
something interesting happens:  effects of a finite cutoff show up even
at the classical level (i.e. before doing the functional integral over
X(z)).  To see this, consider the second order term in the
proper time equation:
\be                     \label{eq:IV.1}
<e^{ik'X(z)}
\int _{2a}^{z-a}du A_{\mu}[X(u)]\p _{u}X^{\mu}(u)
\int _{a}^{u-a}dw A_{\nu}[X(w)]\p _{w}X^{\nu}(w)
e^{ikX(0)}>
\ee
When we perform a gauge variation we get:
\be                 \label{eq:IV.2}
<e^{ik'X(z)}
\Lambda [X(z-a)]\int _{a}^{z-2a} dw A_{\nu}[X(w)]\p _{w}X^{\nu}(w)
e^{ikX(0)}>
\ee
\[
-<e^{ik'X(z)}
\int _{2a}^{z-a}du A_{\nu}[X(u)]\p _{u}X^{\nu}(u) \Lambda [X(a)]
e^{ikX(0)}>
\]
\[
-<e^{ik'X(z)}
\int _{2a}^{z-a} du  \Lambda [X(u)] A_{\nu}[X(u-a)]\p _{u}X^{\nu}(u-a)
e^{ikX(0)}>
\]
\[
+<e^{ik'X(z)}  \int _{2a} ^{z-a} du
 A_{\nu}[X(u)]\p _{u}X^{\nu}(u) \Lambda [X(u-a)]
e^{ikX(0)}>.
\]
Note that we have consistently imposed the rule that `$a$' is
the distance of closest approach between two vertex operators
by appropriate choice of the limits of integration in (\ref{eq:IV.2}).
Notice that the first two terms in expression (\ref{eq:IV.2})
can be cancelled in the
usual way by a variation $\delta \phi = -i \Lambda \phi $
on the first order
term considered in the previous section.  The other two terms, however,
remain to be cancelled.
The third and fourth terms in (\ref{eq:IV.2}) would cancel if`$a$'
were equal to zero.  Their sum is thus proportional to `$a$'.
and can be written
as the sum of three terms:
\be               \label{eq:IV.3}
\int_{a}^{z-a} du A_{\nu}[X(u)]\p _{u}X^{\nu}(u)
[\Lambda [X(u-a)] -
\Lambda [X(u+a)]]
\ee
\[
-\int_{a}^{2a} du A_{\nu}[X(u)]\p _{u}X^{\nu}(u)
[\Lambda [X(u-a)]
\]
\[
+\int_{z-2a}^{z-a} du
[\Lambda [X(u+a)]
A_{\nu}[X(u)]\p _{u}X^{\nu}(u)
\]
The first term is what we get by modifying the limits
of integration of the third and fourth terms in (\ref{eq:IV.2})
so that they both go from $a$ to $z-a$.  The remaining terms (of
(\ref{eq:IV.3})) compensate the $O(a)$ errors that arise from
such a modification (of limits).
Each of the terms in (\ref{eq:IV.3})
can, in turn, be expanded in powers of `$a$'.
The lowest order terms are :
\be
\int_{a}^{z-a} du A_{\nu}[X(u)]\p _{u}X^{\nu}(u) (-2a)
\p _{u} \Lambda [X(u)]  \label{eq:IV.4}
\ee
\[
-aA_{\mu}\p_{u}X^{\mu}(a) \Lambda (a) +a \Lambda (z-a) A_{\mu}
\p_{u}X^{\mu}(z-a)
\]
\be                                    \label{eq:IV.5}
=-2a\int _{a}^{z-a} du A_{\mu}\p _{\nu} \Lambda \p _{u}X^{\mu}
\p _{u}X^{\nu} -a[
A_{\mu}\p_{u}X^{\mu}(a) \Lambda (a) -\Lambda (z-a) A_{\mu}
\p_{u}X^{\mu}(z-a)]
\ee
The first term in (\ref{eq:IV.5})
 is a `bulk' term and the second one is a boundary
term.  The first term, in fact, looks like the vertex operator for a
massive mode.  Thus, consider a massive field $S_{\mu \nu}$ (see eq.
(\ref{eq:IV.8}))
with the transformation law:
\be          \label{eq:IV.6}
\delta S _{\mu \nu} = 2a ( A_{\mu} \nup \Lambda + A_{\nu}\mup \Lambda)
\ee
We could add the background field $a\int S_{\mu \nu} \p _{u} X ^{\mu}
\p _{u} X ^{\nu} du$ to the sigma-model action, (along with $\int
A_{\mu} \p _{u} X^{\mu} du $) and the gauge variation of the first order
contribution to the proper time equation:
\be                         \label{eq:IV.7}
<e^{ik'X(z)} a \int _{a}^{z-a} du S_{\mu \nu}\p _{u} X^{\mu} \p _{u}
X^{\nu} e^{ikX(0)}>
\ee
would cancel the first term in
(\ref{eq:IV.5}).
The boundary term in (\ref{eq:IV.5}) can also be cancelled
as follows:  Consider, again, the second mass level in string theory.
In addition to $S_{\mu \nu}$, there is an auxiliary
field $S_{\mu}$, and the complete vertex operator is [14]:
\be          \label{eq:IV.8}
I=a \int dz \frac{1}{2}S_{\mu \nu} \p X^{\mu} \p X^{\nu} +
S_{\mu}\pp X^{\mu}
\ee
and the usual `free' gauge transformation of string theory is [14]:
\be
\delta S_{\mu \nu} = \mup \Lambda _{\nu} + \nup \Lambda _{\mu}\,\,\;
\delta S_{\mu} = \Lambda _{\mu}
\ee
Under these transformations
\be
\delta I = a \int _{c}^{d}dz [\p  \Lambda _{\nu} \p  X^{\nu}
+ \Lambda _{\nu} \pp X^{\nu} ]   \: = \: a\int _{c}^{d}
dz \p _{z}[\Lambda _{\nu} \p _{z} X^{\nu}]
\ee
\[
= \: a(\Lambda _{\nu} \p _{z}X^{\nu} (d) -
\Lambda _{\nu} \p _{z}X^{\nu} (c))
\]
Clearly if we choose \be \Lambda _{\nu} = - A_{\nu} \Lambda \ee
we can cancel the second term in (\ref{eq:IV.5}).
Thus, we conclude that, at least
classically, and to lowest order, gauge invariance can be restored by
the addition of massive modes with an appropriate gauge transformation
law.
In fact, one can even say, based on the above, that imposing the
U(1) gauge invariance associated with the {\em massless} vector
, on a theory with {\em finite} world sheet cutoff requires
the presence of {\em massive} fields.  We find this very
interesting.

Now, the validity of the Taylor expansion, when quantum mechanics
is turned
on, needs to be checked.  We go back to (\ref{eq:IV.3})
and consider the first
term.  Perform an operator product expansion on each piece separately:
\begin{eqnarray}
\lefteqn{A_{\mu}(p):e^{ipX(u)}\p _{u}X^{\mu}(u):\Lambda(q)
:e^{iqX(u-a)}:= }  \\  &  &
 A_{\mu}(p) \Lambda (q) :(\frac{iq^{\mu}}{a} + \p _{u} X^{\mu})
\mid a \mid ^{p.q}e^{i(p+q)X(u)}(1-iaq.\p _{u} X + ...):
\nonumber
\end{eqnarray}
\begin{eqnarray}
\lefteqn{A_{\mu}(p):e^{ipX(u)}\p _{u}X^{\mu}(u):\Lambda(q)
:e^{iqX(u+a)}:= }  \\  &  &
A_{\mu}(p) \Lambda (q) :(\frac{iq^{\mu}}{-a} + \p _{u} X^{\mu})
\mid a \mid ^{p.q}e^{i(p+q)X(u)}(1+iaq.\p _{u} X + ...):
\nonumber
\end{eqnarray}
Subtracting (4.13) from (4.12), we get for the first term in
(4.3):
\be
 A_{\mu}(p) \Lambda (q) :(\frac{2iq^{\mu}}{a} -2iaq.\p _{u} X
  \p _{u} X^{\mu})
\mid a \mid ^{p.q}e^{i(p+q)X(u)}:
\ee
This corresponds to the operator
\be     \label{eq:IV.9}
[-2a:\nup \Lambda A_{\mu}\p _{u} X^{\mu} \p _{u} X^{\nu}: +
\frac{2}{a}: A^{\mu} \mup \Lambda :] \mid a \mid ^{p.q}
\ee
(If we assume that a normal ordering factor $a^{\frac{p^{2}}{2}}$
accompanies each exponential $e^{ipX}$, then, the factor
$a^{pq}$ in (\ref{eq:IV.9})
is part of the normal ordering associated with
$e^{i(p+q)X}$).
The first term in (\ref{eq:IV.9})
is what we got by the Taylor expansion in eqn.
(\ref{eq:IV.5}).
The second term is a normal ordering effect - it is a contraction
of the two $ \p _{u} X $'s.  It corresponds to a tachyon like
operator and thus represents a gauge transformation on the background
tachyon field :
\be        \label{eq:IV.10}
\delta \phi _{T}= -2 A_{\mu} \mup \Lambda
\ee
We can do a similar analysis for the boundary terms and one finds
that in addition to the terms in (\ref{eq:IV.5})
one needs a term of the form:
\be
\int du \p _{u} (A_{\mu} \mup \Lambda )= \int du \nup (A_{\mu}
\mup \Lambda ) \p _{u} X ^{\nu}
\ee
This is clearly a U(1) gauge transformation of $A_
{\mu}$ itself with parameter $A_{\mu} \mup \Lambda $.
We have thus shown that the boundary terms can be compensated
by usual linear gauge transformations of $S_{\mu \nu}
, S_{\mu} and A_{\mu}$ merely by redefining the gauge parameter.
\footnote {It should be pointed out, that an alternative way is to modify
the transformation of $\phi$.  If we add a piece
$\delta \phi = i A_{\mu} \mup \Lambda \phi $ it would do
just as well.}
Thus we conclude that
(to this order) the theory is gauge invariant with a finite cutoff
provided we modify the gauge transformations of the massive spin 2,
the vector,
and of the tachyon in well defined ways.

There is one consistency check that can be quickly performed.
One can check
whether the non-linear pieces introduced in the gauge transformation law
of $S_{\mu \nu} , \phi _{T} $ are consistent with their interacting
equations of motion.  Obtaining the fully covariant interacting equations
of motion is a complicated story.  This has been done in some
detail in \cite{AC}, in the case where the electromagnetic field
is slowly varying in space-time, which is all we need to lowest order.
One can  also write down the leading
pieces (lowest order in momentum) just by considering the OPE of $A_{\mu}
\p _{z} X^{\mu}$ with itself:
\begin{eqnarray}
\lefteqn{
A_{\mu} (p) :
\p_ {z} X^{\mu} (z)
e^{ipX(z)} :
A_{\nu} (q) :
\p_ {z} X^{\nu} (z+a)
e^{iqX(z+a)} : = } \nonumber   \\
& & A_{\mu} (p)
A_{\nu} (q) :
\p_ {z} X^{\mu} (z)
\p_ {z} X^{\nu} (z+a)
e^{ipX(z)+iqX(z+a)} : \mid a \mid ^{p.q} + \nonumber  \\
& & A.A \frac{1}{a^{2}}
:e^{ipX(z)+iqX(z+a)} : \mid a \mid ^{p.q} + higher \, order \, in \, p,q
\end{eqnarray}
Thus to lowest order in momentum:
\be
=[A_{\mu} (x)A_{\nu}(x) \p_{z} X^{\mu} \p_{z}X^{\nu} + O(a) ]
+\frac{A(x).A(x)}{a^{2}}  + O(a)
\ee
Thus the equation of motion of $S_{\mu \nu}$, which starts out as $(\frac
{p^{2}}{2} -1)S_{\mu \nu}$ gets modifies to
\be      \label{eq:IV.S}
(\frac{p^{2}}{2}-1)S_{\mu \nu} + 2 A_{\mu} A_{\nu} \approx
-S_{\mu \nu} + 2 A_{\mu} A_{\nu} \approx = 0
\ee
and the equation for $\phi$ becomes :
\be                     \label{eq:IV.A}
(\frac{p^{2}}{2}+1)\phi +  A_{\mu} A^{\mu} \approx
\phi + A.A \approx 0
\ee
Both these, (\ref{eq:IV.S}) and (\ref{eq:IV.A})
are consistent with the modifications
(\ref{eq:IV.6}) and (\ref{eq:IV.10}) respectively.

Thus we conclude that keeping a finite cutoff while retaining
gauge invariance forces one to introduce background massive modes,
and to modify  their transformation laws.  These modifications
are consistent with what one expects from their interacting
equations.  We expect that at higher order in `$a$' one will need other
massive modes as well.  Now, all this is not surprising.  A string
is a non local object, (but with local interactions)
which is why we have an infinite number of point
particles.  A finite cutoff makes the theory non-local, albeit
in a crude way.  Requiring gauge invariance is a way to make this
more refined and `string'-like.  This requires massive modes.  One can
also argue, as in the introduction, from the viewpoint of the
renormalization group, that keeping a finite cutoff entails retaining
all the irrelevant operators that correspond to massive modes.
Thus there
are different ways to understand or rationalize these results.
Nevertheless
we think that deducing the existence of massive modes
and their transformation properties
from the requirement
of ordinary (`massless') gauge invariance, is very interesting.
\newpage
\section{Conclusions}

In this paper we have investigated two interrelated topics
i)gauge invariance
at the interacting level , ii) keeping a finite cutoff and
going off-shell.
We have a technique for dealing with gauge particles in the proper-time
framework.  This was an extension to vector-vector interactions
of the results of ref[15] for free gauge particles.  We do not yet
know
how to do extend this to an off shell calculation.  In section 3
we discussed
the simpler version of the above problem: going off-shell
with a finite cutoff in the case of the covariant Klein Gordon equation.
We presented one possible form of the interacting term that satisfies
some basic properties of gauge invariance and of having the right
on-shell limit.  There are other solutions possible.  In particular,
if one does the same calculation, but on the boundary of a disc, one
will have manifest cyclic symmetry.  In order to proceed further, one
needs a prescription for going from
3-point functions to 4-point functions or higher n-point
functions.  This, we think, is the most pressing issue in this
approach.
In sec. 4, in studying the 4-point function directly, we discovered
that a non-zero cutoff, along with the requirement of gauge invariance,
predicts not only the existence of masssive modes,
 but also the right transformation law.  We find this promising.
It would be interesting to extend these results to all the massive
modes and higher invariances.  Finally, on a more speculative note, the
idea of a finite world sheet cutoff has to get translated to a finite
space-time cutoff.
\newpage
\appendix
\section{Appendix}
\setcounter{equation}{0}
We will prove eqn.(\ref{eq:Id})
by proving the following relation:
\be
\frac{(i\ko .X)^{n}}{n!}i\ki \p X =
\frac{1}{(n+1)!} \p ((i\ko X)^{n} i\ki .X)
+\frac{n}{(n+1)!}((i\ko X)^{n-1} X^{\sigma} \p X^{\mu}  (i)^{2}
\ko ^{[ \sigma} \ki ^{\mu ]} )
\ee
L.H.S. of (A.1) obviously sums to $e^{i \ko X} i\ki \p X$. The
second term is
\be
\frac{1}{(n+1)}
\p (\frac{(i\ko X)^{n}}{n!} i\ki .X) =
\int _{0}^{1}
d \al \al ^{n}
\p (\frac{(i\ko X)^{n}}{n!} i\ki .X) =
\ee
\[
\int _{0}^{1}
d \al
\p (\frac{(i\al \ko X)^{n}}{n!} i\ki .X)
\]
which sums to
$\int _{0}^{1}
d \al
\p (e^{i \alpha \ko X )}i \ki X)$.  Similarly the last term becomes
\be
\sum _{n=1}^{\infty}
\frac{n}{(n+1)!}((i\ko X)^{n-1} X^{\sigma} \p X^{\mu}  (i)^{2}
\ko ^{[ \sigma} \ki ^{\mu ]} )
=
\ee
\[
=
\int _{0}^{1}  d \al
\sum _{n=1}^{\infty}
\frac{1}{(n-1)!}((i\al \ko X)^{n-1}
X^{\sigma} \p X^{\mu}  (i)^{2}
\al \ko ^{[ \sigma} \ki ^{\mu ]} )
\]
\[
=
\int _{0}^{1} d \al
e^{i \al \ko X}
X^{\sigma} \p X^{\mu}  (i)^{2}
\al \ko ^{[ \sigma} \ki ^{\mu ]} )
\]
Thus we have to prove (A.1), which we shall do by recursion:
Consider the relation
\be
\Xmi \Xmt ...\Xmnm \p \Xmn =
\frac{1}{n}
\p(\Xmi \Xmt ...\Xmn ) +
\ee
\[
\frac{1}{n}
\underbrace{
[\Xmi \Xmt ... X^{[ \mnm } \p X^{\mn ]} +
\Xmnm \Xmi \Xmt ...X^{[ \mu _{n-2}} \p X^{\mn ]} +....]}_{n-1 \, cyclic
\, permutations}
\]
Multiply by
$(i)^{n}\ko ^{\mi} \ko ^{\mt} ...\ko ^{\mnm} \ki ^{\mn}$ to get
\be
(i\ko X)^{n-1} (i\ki \p X) =
\frac{1}{n}
\p
((i\ko X)^{n-1} i\ki  X) +
\ee
\[
\frac{1}{n}
\underbrace{
[(i\ko X)^{n-2}X^{[ \mnm } \p X^{\mn ]}
i^{2} \ko ^{\mnm} \ki ^{\mn}+
[(i\ko X)^{n-2}
X^{[ \mu _{n-2}} \p X^{\mn ]}
i^{2} \ko ^{\mu _{n-2}} \ki ^{\mn}+ ...]}_{n-1 \, terms}
\]
\be
\Rightarrow
(i\ko X)^{n-1} (i\ki \p X) =
\frac{1}{n}
\p
((i\ko X)^{n-1} i\ki  X) +
\ee
\[
\frac{n-1}{n}
(i\ko X )^{n-2} X^{\sigma }\p X^{\mu} \ko ^{[ \sigma} \ki ^{\mu ]}
i^{2}
\]
Dividing by $(n-1)!$ and replacing $n$ by $n+1$ gives (A.1). Thus it
remains to prove (A.4).

We have
\be
\Xmi \Xmt ... \Xmn \p \Xmnp =
\ee
\[
\Xmi \frac{1}{n}
\underbrace{[
\p (\Xmt ...\Xmnp ) +
\Xmt X^{\mu _{3}}...X^{[ \mu _{n}}\p X ^{\mu _{n+1} ]}
+ \Xmn \Xmt ...X^{[ \mu _{n-1}}\p X^{\mnp ]} +...] }_{n-1 \, cyclic \,
permutations}
\]
where we have used (A.4).
\be
=
\frac{1}{n}
1/2(\Xmi \p (\Xmt ...\Xmnp ) +
\p \Xmi (\Xmt ...\Xmnp )) +
\ee
\[
1/2(\Xmi \p (\Xmt ...\Xmnp ) -
\p \Xmi (\Xmt ...\Xmnp )) +
\]
\[
\frac{1}{n}
[\Xmi \underbrace{
\Xmt X^{\mu _{3}}....X^{[ \mn }}_{n-1 \, perm. \, of \,  \mt ...\mn}
\p X^{\mnp ]}]
\]
\be
=
\frac{1}{n}
[
1/2 \p (\Xmi ....\Xmnp )
+1/2(\Xmi \Xmt \p X^{\mu _{3}}...\Xmnp +.....
+\Xmi.....\Xmn \p \Xmnp ) ]+
\ee
\[
\frac{1}{n}
[\Xmi \underbrace{
\Xmt X^{\mu _{3}}....X^{[ \mn }}_{n-1 \, perm. \, of \,  \mt ...\mn}
\p X^{\mnp ]}]
\]
In going from (A.8) to (A.9) we have dropped a term manifestly
antisymmetric in $\mi \mt $ since the LHS is manifestly
symmetric.  From (A.7) and (A.9) we have:
\be
(1- \frac {1}{2n})
\Xmi \Xmt ... \Xmn \p \Xmnp =
\frac{1}{n}
[1/2 \p (\Xmi ....\Xmnp )+
\ee
\[
+1/2(\Xmi \Xmt \p X^{\mu _{3}}...\Xmnp +.....
+\Xmi...\p \Xmn  \Xmnp ) ]+
\]
\[
+\frac{1}{n}
[\Xmi \underbrace{
\Xmt X^{\mu _{3}}....X^{[ \mn }}_{n-1 \, perm. \, of \,  \mt ...\mn}
\p X^{\mnp ]}]
\]
We can now add all cyclic permutations of $\mi ....\mn $ .  The LHS
is manifestly symmetric and we get a factor of $n$. The resultant
equation is:(We have divided (A.10) by $\frac{2n-1}{2n}$)
\be
n
\Xmi \Xmt ... \Xmn \p \Xmnp =
\frac{n}{(2n-1)}\p (\Xmi ....\Xmnp )+
\ee
\[
\frac{n-2}{2n-1}
\p(\Xmi...\Xmn) \Xmnp
\]
\[
+
\frac{2(n-1)}{2n-1}[\underbrace{\Xmi ....X^{[ \mn} \p X^{\mnp ]}}_{n
\, perm }]
\]
Combining terms we get
\be
(2n^{2} -2)
\Xmi \Xmt ... \Xmn \p \Xmnp =
2(n-1)
\p (\Xmi ....\Xmnp )+
\ee
\[
2(n-1)[\underbrace{\Xmi ....X^{[ \mn} \p X^{\mnp ]}}_{n
\, perm }]
\]
Dividing throughout by $2n^{2}-2$ gives us a relation that is the
same as (A4) with $n$ replaced by $n+1$.  This proves the
recursion.  Since $X^{\rho} \p X^{\mu} = 1/2 \p (X^{\rho} X^{\mu})
+1/2 X^{[ \rho} \p X^{\mu ]} $, the relation is true for $n=2$ and
this completes the proof.

\end{document}